\documentclass[aps,prb,twocolumn,superscriptaddress,showpacs]{revtex4}
\usepackage{graphicx}
\usepackage{hyperref}
\usepackage{color}
\usepackage{soul}

\begin{document}

\title{Simple model for the vibrations of embedded elastically cubic
nanocrystals}
\author{Lucien Saviot}
\email{lucien.saviot@u-bourgogne.fr}
\affiliation{Laboratoire Interdisciplinaire Carnot de Bourgogne,
UMR 5209 CNRS-Universit\'e de Bourgogne,
9 Av. A. Savary, BP 47 870, F-21078 Dijon Cedex, France}

\author{Daniel B. Murray}
\email{daniel.murray@ubc.ca}
\affiliation{Department of Physics, University of British Columbia Okanagan,
3333 University Way, Kelowna, British Columbia, Canada V1V 1V7}

\author{Eug\`ene Duval}
\author{Alain Mermet}
\author{Sergey Sirotkin}
\affiliation{Laboratoire de Physico-Chimie des Mat\'eriaux Luminescents,
Universit\'e de Lyon, Universit\'e Claude Bernard Lyon 1, UMR 5620 CNRS,
69622 Villeurbanne, France}

\author{Mar\'ia del Carmen Marco de Lucas}
\affiliation{Laboratoire Interdisciplinaire Carnot de Bourgogne,
UMR 5209 CNRS-Universit\'e de Bourgogne,
9 Av. A. Savary, BP 47 870, F-21078 Dijon Cedex, France}

\begin{abstract}
The purpose of this work is to calculate the vibrational modes of an
elastically anisotropic sphere embedded in an isotropic matrix. This has
important application to understanding the spectra of low-frequency Raman
scattering from nanoparticles embedded in a glass matrix. First some low
frequency vibrational modes of a free cubically elastic sphere are found to
be nearly independent of one combination of elastic constants. This is then
exploited to obtain an isotropic approximation for these modes which
enables to take into account the surrounding isotropic matrix. This method
is then used to quantatively explain recent spectra of gold and copper
nanocrystals in glasses.
\end{abstract}

\maketitle

\section{Introduction}

Low-frequency inelastic light scattering by metallic nanoparticles, which
is due to their mechanical vibrations, has been the focus of attention of
many researchers during the last thirty
years.\cite{GerstenPRL80,Fujii91,erratumFujii91,BachelierPRB04} This
scattering is similar to surface enhanced Raman scattering from molecules
close to such a metallic nanostructure \cite{GerstenPRL80} making it an
interesting complementary way to study this complex phenomenon having many
applications for very sensitive detection. The enhancement due to using
laser excitations resonant with the dipolar plasmon in such nanostructures
and the high quality samples available today are responsible for
low-frequency Raman spectra having an unmatched number of features
\cite{AdichtchevPRB09} compared to non-metallic nanoparticles.

The interpretation of such spectra is very challenging as many
parameters have to be taken into account at the same time. The
mechanical vibrations depend on the shape of the nanoparticles
but also on their inner structure and on the surrounding medium.
In this work, we report on a new approach enabling taking into
account all these parameters for a spherical nanoparticle
having a cubic lattice embedded in a glass matrix.

The vibrational modes of an elastically isotropic sphere which is
free\cite{lamb1882} or embedded in an infinite isotropic
matrix\cite{dubrovskiy81,MurrayPRB04} are known exactly. When the elastic
constants are not isotropic, a numerical approach such as the one known as
Resonant UltraSound (RUS)\cite{visscher} can find mode frequencies and
displacement fields of free
nanoparticles.\cite{mochizukiJAP88,OdaGJI94,SaviotPRB09} What has been
missing up until now is a description of the vibrations of an elastically
anisotropic sphere embedded in an isotropic matrix. The case of
anisotropic elasticity has been discussed for the problem of the scattering
of acoustic waves \cite{HasheminejadAP08,ZuckermanPRB08} but none of these
approaches provides the eigendisplacements required for modeling the
coupling of the vibrations with electrons which is at work in all the
optical techniques used to detect such vibrations.

Current experimental results such as those presented in
Ref.~\onlinecite{AdichtchevPRB09} have already shown the need for
a model without any of these limitations. The interpretation in
that paper took into account the elastic anisotropy to
qualitatively explain the splitting of the lowest frequency Raman
peaks but it failed to provide a quantitative description due to
the significant coupling with the embedding matrix. The present
work fills that hole by showing that it is possible to choose an
isotropic approximation of the system for the most intense Raman
active vibrations which enables the prediction of the position of
the Raman peaks and provides the vibrational displacement
fields required for the calculation of the Raman intensities.

\section{Method}

Exact solutions for the vibrations of isotropic free spheres can be
classified as spheroidal and torsional and will be noted $S_{\ell,m}^n$ and
$T_{\ell,m}^n$ respectively in the following with $\ell$ and $m$ being the
usual angular momentum and its $z$-component and $n$ being an index used to
label the eigenmodes by increasing frequency starting from $n=1$ as in a
previous work.\cite{SaviotPRB09} For spheres whose diameter is small
compared to the wavelength of light, the Raman-active vibrations are $S_0$
and $S_2$ (for every $m$ and $n$).\cite{duval92} In the present work, RUS
calculations have been used to model the vibrations of elastically
anisotropic spheres as in a previous work \cite{SaviotPRB09} by expanding
the displacements onto $x^iy^jz^k$ functions with $i+j+k \le 20$.

The approach of this paper was inspired by 
the calculations for elastically anisotropic cuboctaedra
in Ref.~\onlinecite{StephanidisPRB07}.
The idea was to calculate mode frequencies using only the
speed of sound along a single propagation direction.
The 5-fold degenerate $S_2^1$ modes of an isotropic sphere
are split by cubic elasticity 
into two degenerate modes with $E_g$ symmetry and three
with $T_{2g}$ symmetry ($O_h$ point group).
Due to the symmetry of the displacements
of the $E_g$ and $T_{2g}$ vibrations, their
frequencies were approximated using sound speeds in
particular directions instead of 3D-averaged ones.
It should be noted that all the $E_g$ and $T_{2g}$ vibrations (as well as
the $A_{1g}$ vibrations) are Raman
active but only the ones sharing a strong similarity with the $S_2$ modes
are expected to contribute significantly to the Raman spectra due to the
surface deformation scattering mechanism.\cite{BachelierPRB04}

In order to confirm the
validity of this simple approach for spherical nanoparticles, we use a
method similar to the one used in a previous work.\cite{SaviotPRB05}
Studying the frequency changes resulting from a continuous variation of the
elasticity of the material the sphere is made of provides some insight into
the nature of the vibrations. A cubic material has three independent  elastic constants
$C_{11}$, $C_{12}$ and $C_{44}$ instead of two for isotropic
elasticity for which $C_{44}=\frac{C_{11}-C_{12}}{2}$.
We consider the case of gold nanoparticles
because of available inelastic light scattering experimental data to
compare with and also because gold has a very strong elastic anisotropy
making it a good system to test the validity of the isotropic
approximations. We use the following parameters: $C_{11}=191$~GPa,
$C_{12}=162$~GPa and $C_{44}=42.4$~GPa and mass density
$\rho=19.283$~g.cm$^{-3}$. Since we are mainly interested in Raman active
vibrations, we will focus on the vibrations coming from the isotropic
$S_2^1$ mode which are the main features in the low-frequency Raman spectra
of gold nanoparticles.

\begin{figure}
\includegraphics[width=\columnwidth]{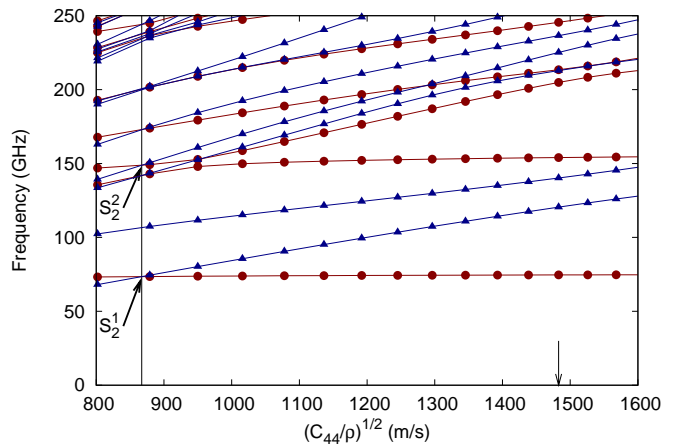}
\caption{\label{AuC44}(Color online) Variation of the eigenfrequencies
of the lowest frequency $E_g$ (circles, red online) and $T_{2g}$
(triangles, blue online) vibrations of a gold sphere
of radius 5~nm when varying $C_{44}$. The vertical arrow indicates the real
abscissa for gold and the vertical line corresponds to an isotropic system.}
\end{figure}

The potential for a great simplification in handling these modes, at least
in some cases, can be clearly seen as follows. We plot eigenfrequencies
for a gold sphere as a function of $C_{44}$ in Fig.~\ref{AuC44}, instead
of leaving $C_{44}$ fixed at its normal value for gold. The two lowest
frequency $E_g$ branches, i.e. the fundamental mode and first overtone, are very nearly flat. The frequency change for the
lowest $E_g$ branch is 2\% while $C_{44}$ is multiplied by 4. This
indicates that the corresponding modes approximately do not depend on
$C_{44}$ or the associated transverse sound speed. This approximate
flatness is essential to our approach. In practice, it only holds very well
for the lowest frequency modes, and successively less well for higher
frequency. The projection of the displacements\cite{SaviotPRB09} of the
two lowest frequency $E_g$ vibrations obtained for
$\sqrt{C_{44}/\rho}=1600$~m/s onto those obtained for
$\sqrt{C_{44}/\rho}=800$~m/s is very close to 1
($\simeq 0.9968$). This demonstrates that the displacement field of these modes do
not significantly change with $C_{44}$ either. Since the mode
approximately does not depend on $C_{44}$, we are free to arbitrarily
change $C_{44}$ to a different value which is convenient for us.
Specifically, we can always choose $C_{44}$ to change the gold into an
isotropic material, \textit{i.e.} $C_{44}=\frac{C_{11}-C_{12}}{2}$.

\begin{figure}
\includegraphics[width=\columnwidth]{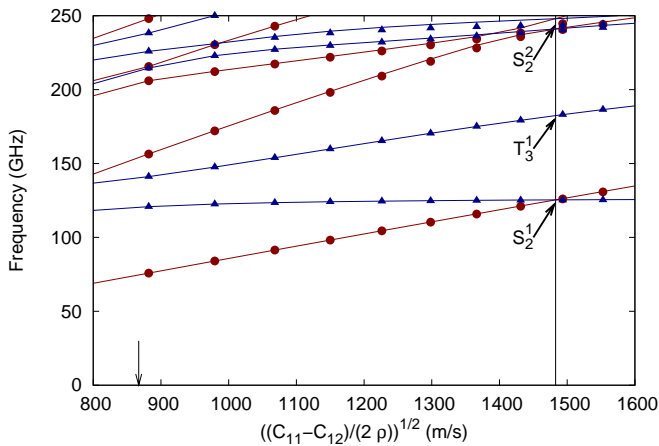}
\caption{\label{AuC1112}(Color online) Variation of the eigenfrequencies
of the $E_g$ (circles for varying $C_{12}$ and lines going through the
circles for varying $C_{11}$, red online) and $T_{2g}$ (triangles for
varying $C_{12}$ and lines going through the triangles for varying
$C_{11}$, blue online) vibrations of a gold sphere of radius 5~nm as
a function of $C_{11}-C_{12}$. The vertical arrow indicates the real
abscissa for gold and the vertical line corresponds to an isotropic
system.}
\end{figure}

The lowest frequency $T_{2g}$ modes depend on $C_{44}$ as can be seen by
their frequency variations in Fig.~\ref{AuC44}.  We also consider the
variation of their frequencies as a function of $C_{11}-C_{12}$ in
Fig.~\ref{AuC1112} while varying either $C_{11}$ or $C_{12}$ one at a time.
There is nearly perfect agreement in this figure between varying $C_{11}$
or $C_{12}$.  It is also very good for vibrations having other irreducible
representations (not shown).  This demonstrates that $C_{11}-C_{12}$ is a
good choice for a parameter rather than $C_{11}$ only or $C_{12}$ only. Furthermore,
the lowest frequency $T_{2g}$ vibrations which come from the $S_2^1$ modes
do not depend on the corresponding transverse sound speed
$\sqrt{\frac{C_{11}-C_{12}}{2 \rho}}$ as can be seen by the almost flat
variation. The small variation observed near the value of gold is
due to the anti-crossing between the branches of the two lowest frequency
$T_{2g}$ branches. While the lowest branch is associated with the $S_2^1$
mode, the next upper one comes from $T_3^1$ modes. Had a less anisotropic
material been chosen, this anti-crossing pattern would have been less
pronounced.  Still, as will be discussed later, neglecting the mixings
between these branches is a reasonable choice in many cases.

In both cases, the mode-dependent isotropic
approximation is obtained by satisfying
$C_{44}=\frac{C_{11}-C_{12}}{2}$. The displacements associated with the
mode-dependent isotropic approximations can be constructed using symmetry
arguments \cite{lacroix,OdaGJI94} from the $S_{2,m}$ displacements. $S_{2,0}$ and
$\frac{S_{2,2}+S_{2,-2}}{\sqrt{2}}$ are two orthonormal $E_g$ modes and
$\frac{S_{2,2}-S_{2,-2}}{\sqrt{2}}$, $S_{2,1}$ and $S_{2,-1}$ are the three
$T_{2g}$ orthonormal modes.
As has been shown\cite{SaviotPRB05}, the key parameter for the calculation of the $S_2^1$ modes
is
the transverse isotropic sound speed although the longitudinal one has also
a very small contribution.
Choosing the
isotropic longitudinal sound speed is not critical also due to
the fact that the quasi-longitudinal sound speed does not vary much with
the propagation direction in gold. As a result, we simply used the
3D-averaged longitudinal sound speed as in previous works. This value is
also the most appropriate one for an isotropic approximation of the
breathing modes $S_0$ as has been already shown.\cite{SaviotPRB09}

\section{Application}

\subsection{Free gold nanocrystals}

\begin{figure}
\includegraphics[width=\columnwidth]{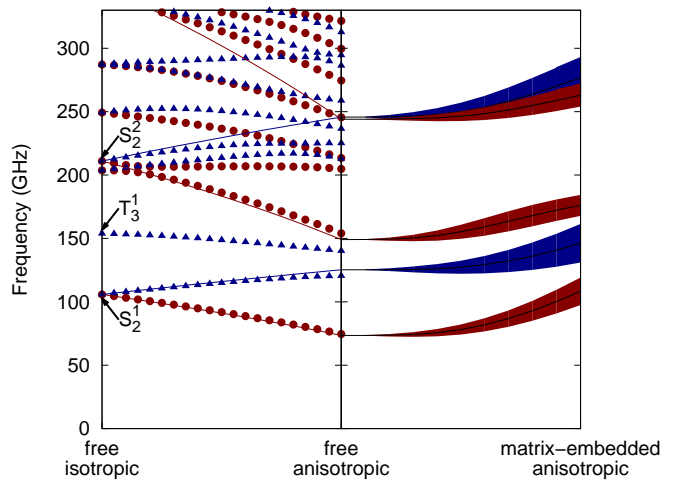}
\caption{\label{AuISOAni}(Color online) Left: variation of the
frequencies using RUS (symbols) and Lamb with mode-dependent sound
speeds (lines) when linearly varying the $C_{ij}$ from isotropic
to anisotropic gold. Radius $R=5$~nm. -- Right: Matrix induced
broadening and frequency shift of the same modes. $v_L=\alpha*5020$~m/s,
$v_T=\alpha*3010$~m/s and $\rho=\alpha*2.97$~g/cm$^3$ for the matrix and the
abscissa $\alpha$ is varied from 0 to 1 so as to reach the same value as in
Ref.~\onlinecite{StephanidisPRB07}.}
\end{figure}

Now the calculation of the eigenfrequencies of the modes coming from
the spheroidal quadrupolar vibrations ($S_2^1$) for a sphere made of a
material with cubic elasticity is separated into two isotropic problems
which can each be solved exactly. The result of such calculations for
a gold sphere are presented in the left part of Fig.~\ref{AuISOAni}
together with RUS calculations for varying anisotropy by using
$C_{ij}(x) = C_{ij}^{iso} + x * (C_{ij}^{ani}-C_{ij}^{iso})$ with
$0 \le x \le 1$, $x$ being the abscissa, $C^{iso}$ being the isotropic
gold stiffness tensor obtained from the 3D-averaged sound speeds and
$C^{ani}$ being the anisotropic one. The agreement between both kinds
of calculations for the lowest three approximated branches is very
good.
The small deviation from the lowest frequency $T_{2g}$ branch close to the
``free anistropic'' limit can be
attributed to the fact that the mixing with the next $T_{2g}$ branch
coming from $T_3^1$ is taken into account only in the RUS calculation.
This mixing has not yet been observed experimentally (it should
manifest as a splitting and intensity sharing between both branches). But
even in that case, the frequency provided by the modified Lamb approach
can be seen as a good approximation of the position of the expected
Raman peak. As a result, this approach provides a simpler description of
the vibrations which is quite suitable to interpret all the currently
available experimental results for free nanocrystals.

\subsection{Matrix embedded nanocrystals}

Since the coupling with an embedding matrix can be taken into account for a
system having spherical symmetry, the isotropic approximations presented
before for free nanocrystals can be extended to calculate the broadening
and frequency shifts due to a surrounding matrix. In the following, we
focus on calculations using the pseudo-modes obtained with the complex
frequency model (CFM).\cite{SaviotPRB09} Other models such as the
core-shell model (CSM)\cite{SaviotPRB09} would be more suitable for the
calculation of Raman spectra but won't be considered here as we focus only
on the positions of the Raman bands. Details about the validity of the CFM
approach using the isotropic approximations are presented in
Appendix~\ref{AppCFM}.

\begin{figure}
\includegraphics[width=\columnwidth]{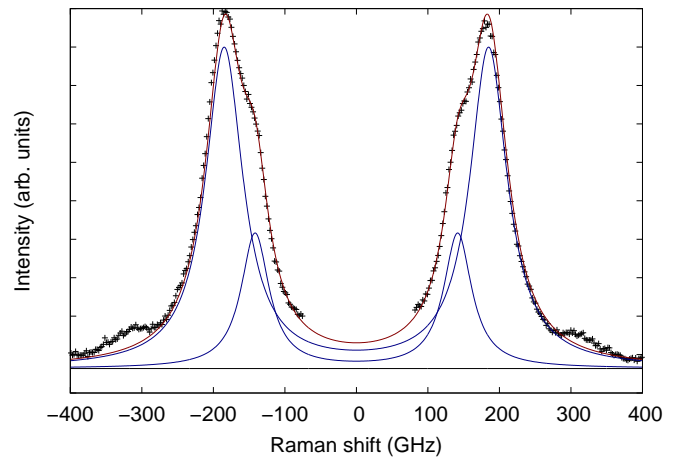}
\caption{\label{fitAu}(Color online) Low-frequency Raman spectrum for
matrix embedded gold nanoparticles (crosses). The fit of the intense
lowest frequency band by two lorentzians and a constant background is
shown with lines.}
\end{figure}

CFM calculations are presented in Fig.~\ref{AuISOAni} and compared to
the experimental spectra of gold nanocrystals in Fig.~\ref{fitAu} and
Ref.~\onlinecite{StephanidisPRB07}.
Fig.~\ref{fitAu} shows the Raman spectra of matrix-embedded gold
nanocrystals. Details about the sample preparation and the spectra
acquisition are the same as those presented in
Ref.~\onlinecite{StephanidisPRB07} with a 64 hour annealing at
$T=455^\circ$C. Fig.~\ref{fitAu} focuses on the intense lowest frequency
peak which has been decomposed into a low frequency $E_g$ lorentzian
(position $141.6 \pm 0.7$~GHz, full width at half maximum (FWHM) $47.5 \pm
1.9$~GHz) and a higher frequency $T_{2g}$ one (position $184.9\pm0.4$~GHz,
FWHM $63.7 \pm 1.2$~GHz).

Fig.~\ref{AuISOAni} (right) presents the evolution from free to
matrix-embedded gold nanocrystals by varying the parameters describing the
matrix (mass density and longitudinal and transverse sound speeds). As
discussed before, the calculations are expected to be accurate only for the
lowest $E_g$ and $T_{2g}$ branches but higher frequency branches are shown
as well for completeness. We used the parameters for the matrix which were
measured by Brillouin scattering in previous
works.\cite{StephanidisPRB07,AdichtchevPRB09} On the right-hand side of
Fig.~\ref{AuISOAni}, the frequency of the $E_g$ branch reaches 108.3~GHz
(FWHM 21.7~GHz) while the $T_{2g}$ one reaches 146.2~GHz (FWHM 30.4~GHz).
While going from free to matrix-embedded nanoparticles, the ratio of the
$E_g$ and $T_{2g}$ frequencies, which does not depend on the size of the
nanocrystals, changes from 0.59 to 0.74. Both values are in very good
agreement with the experimental results presented here (ratio 0.77) and in
Ref.~\onlinecite{StephanidisPRB07} for embedded gold
nanocrystals\cite{StephanidisPRB07} as well as for free gold nanocrystals
(ratio 0.62 in Ref.~\onlinecite{PortalesACSnano10}).  This good agreement
strongly supports the validity of the calculations based on the CFM and
using the isotropic approximations obtained for the free spheres.
Moreover, it enables the accurate determination of the size of the
nanocrystals using the inverse proportionality of the eigenfrequencies with
the diameter of the sphere.

A similar procedure was used to check the validity of our approach with
matrix embedded copper nanoparticles since copper cristallizes in a cubic
structure too.  Red Cu glasses were produced in a similar fashion to the Au
based glasses described in Ref.~\onlinecite{StephanidisPRB07}. By annealing
near the glass transition temperature an initially transparent
sodo-silicate glass containing minute quantities of both Cu$_2$O and SnO,
nanometric clusters of metallic copper are formed. Based on a Maxwell
Garnett description, the volume content of Cu under the form of
nanoparticles is estimated as $10^{-5}$. Similar values are expected for
the previous gold samples. Low-frequency Raman spectra were recorded using
a laser excitation close to the surface plasmon resonance maxima
($\lambda=561$~nm for Cu instead of $\lambda=532$~nm for Au).  Due to the
observed splitting of the lowest frequency band, it is deduced that a
substantial fraction of the formed Cu nanoparticles are monodomain
nanocrystals. The density of the embedding glass is $\rho=2.43$ g.cm$^{-3}$
and the longitudinal and tranverse sound velocities are respectively
$v_L=5800$ m/s and $v_T=3510$ m/s, as determined from Brillouin
spectroscopy. Note that these matrix parameters differ from those of the Au
sample ($\rho=2.97$ g.cm$^{-3}$, $v_L=5020$ m/s and $v_T=3010$ m/s).

\begin{figure}
\includegraphics[width=\columnwidth]{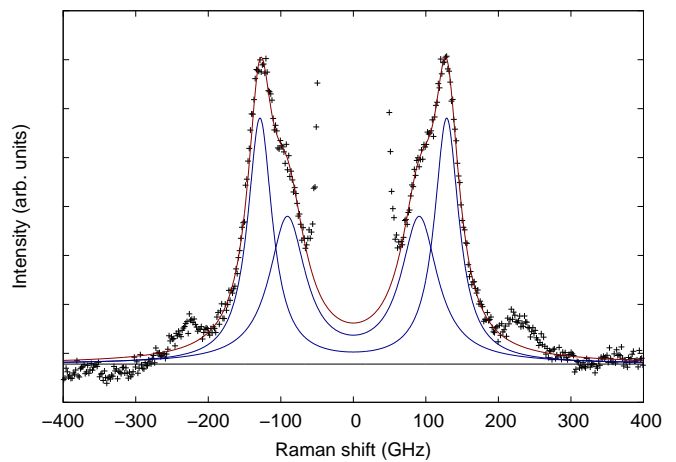}
\caption{\label{fitCu}(Color online) Low-frequency Raman spectrum for
matrix embedded copper nanoparticles (crosses). The fit of the intense
lowest frequency band by two lorentzians and a constant background is
shown with lines.}
\end{figure}

The mass density and $C_{ij}$'s for copper were obtained from
Ref.~\onlinecite{Kittel}. For free monodomain copper nanocrystals, the
calculated ratio of the $E_g$ to $T_{2g}$ frequencies is 0.56 and it
increases to 0.66 when embedded in the corresponding matrix. The ratio
deduced from the fit presented in Fig.~\ref{fitCu} is 0.71.

It should be noted that, as clearly demonstrated
elsewhere\cite{PortalesACSnano10}, multiply-twinned particles also
contribute to the Raman spectra in the same frequency range through a
broader peak. The broadening of the $E_g$ and $T_{2g}$ peaks for matrix
embedded nanocrystals prevents the clear identification of this additional
contribution and therefore it is not possible to reliably fit it. However,
by not taking it into account, the positions of the $E_g$ and $T_{2g}$
peaks are not very accurate which may explain the disagreements between the
calculated and fitted ratios. For the same reason, the fitted intensities
and widths of both peaks are seriously affected by the presence of such a
third contribution. It is therefore important to restrict the usage of the
fitting procedure used in this work to obtain only relatively accurate
positions for the $E_g$ to $T_{2g}$ peaks. These should of course not
depend strongly on the exact shape of the peaks (Lorentzian or Gaussian for
example).

Finally, an additional small peak at higher frequency can be seen in the
spectra presented in Figs~\ref{fitAu} and \ref{fitCu}.  Within the
isotropic approximation, this peak was assigned to the $S_2^2$ degenerated
modes.\cite{AdichtchevPRB09} In the view of Fig.~\ref{AuISOAni},
\textit{i.e.} accounting for the elastic anisotropy and the embedding
matrix, this peak may tentatively be assigned to either the $E_g$ modes
deriving from the $S_2^3$ vibrations or the $T_{2g}$ modes deriving from
the $S_2^2$ modes. Since the latter modes are known to have a very small
surface deformation, their expected Raman scattering cross-section is
expected to be negligible. Therefore we assign this third small peak to the
$E_g$ modes deriving from the $S_2^3$ vibrations.

\section{Conclusion}

We have demonstrated that full anisotropic calculations are often not
required to interpret low-frequency Raman spectra from spherical
nanoparticles with cubic elasticity. Instead, simpler isotropic
calculations with properly chosen sound speeds can provide approximate but
accurate frequencies for the most intense Raman active vibrations and allow
to take into account the medium surrounding the nanocrystals. Calculations
have been successfully compared to experimental results for the case of
monodomain gold and copper nanoparticles in a glass matrix having isotropic
elasticity. While the effect of the elastic anisotropy on the frequencies
of the free vibrations was already known, the present work demonstrates the
measurable impact it also has on the frequency splittings and the
broadenings of the pseudo-mode for matrix embedded nanoparticles. Such is
important for a reliable size evaluation of nanoparticles from their low
frequency Raman spectra.

\begin{acknowledgments}
We acknowledge Serge Etienne for providing some of the samples used in this
work.

Part of his work has been carried out within the FeNoPti$\chi$ project
n$^\circ$ANR-09-NANO-023 funded by the French National Agency (ANR) in
the frame of its 2009 programme in Nanosciences, Nanotechnologies and
Nanosystems (P3N2009)
\end{acknowledgments}

\appendix

\section{Validity of the CFM used with the isotropic approximations}
\label{AppCFM}

Let's consider a gold sphere of radius $R_0$. In order to apply the CFM
approach to the case of spherical nanocrystals having anisotropic
elasticity, we have to check that the displacement fields originally used in
the isotropic CFM case are still approximate solutions of the wave equation
for the anisotropic system for $r<R_0$ provided the correct $C_{ij}$'s are
used.

Without any loss of generality, let's first consider the $E_g$ modes
similar to $S_{2,0}$ ($m=0$). We already know that the isotropic
approximation presented before is valid for the lowest two $E_g$ modes.  We
write the frequencies of these modes as $\nu_1=\frac{A}{R_0}$ and
$\nu_2=\frac{B}{R_0}$ with $A<B$ in the following. This isotropic
approximation is valid whatever the radius of the gold sphere and in
particular for radii $R>R_0$. For a given frequency $\nu<\frac{A}{R_0}$,
let's consider two particular spheres of radii $R_A=\frac{A}{\nu}$ and
$R_B=\frac{B}{\nu}$ so that the first $E_g$ mode of the first particle and
the second $E_g$ mode of the second particle have the same frequency $\nu$.
In the isotropic approximation, these $S_{2,0}$ displacements are linear
combinations of two terms:
$\vec u_L = \vec\nabla j_2(k_L r) P_2(\cos\theta) e^{i \omega t}$
and
$\vec u_T = \vec\nabla \times \vec\nabla \left(\vec r j_2(k_L r)
P_2(\cos\theta)\right)  e^{i \omega t}$
where $j_\ell$ and $P_\ell$ are the
spherical Bessel functions of the first kind and the Legendre polynomials
respectively and $\omega = k_L v_L = k_T v_T$. Since two independent linear combinations of the same fields
($\vec u_L$ and $\vec u_T$) are approximate solutions of the wave equation,
each field is an approximate solution too for $r<R_0$.  Therefore we have
checked that it is possible to apply the CFM approach in the frequency
domain $\nu<\nu_1$.

The matrix often increases the pseudo-mode frequencies up to roughly 50\%
(\textit{i.e.} roughly $\frac{\nu_1+\nu_2}{2}$) in the case of a very hard
and dense matrix. We want to apply the CFM in that frequency range too.
The fact that the calculations are also valid for $R<R_0$ is of no use here
because it only demonstrates that the $S_{2,0}$ displacement field is a
good approximation for the core of the sphere but nothing is known close to
$r=R_0$.  While we can't extend this proof to $\nu>\nu_1$, we can't prove
either that this approximation fails very quickly when increasing $\nu$
above $\nu_1$.
For $T_{2g}$ modes, the situation is also more
problematic because the branch coming from $S_2^2$ mixes significantly with
other $T_{2g}$ branches coming from $S_4^1$, $T_5^1$ and $S_6^1$.
Despite these limitations, applying the CFM can be useful at least as a
rough estimation of the pseudo-modes frequencies.

\bibliography{ani2iso}
\end{document}